\documentstyle[11pt,rafringe,twoside,graphicx,epsfig,makeidx]{article}
\makeindex
\def\edcomment#1{\iffalse\marginpar{\raggedright\sl#1\/}\else\relax\fi}
\reversemarginpar

\begin{document}
\title{Variability and Velocity of Superluminal Sources}
\author{M. H. Cohen, M. A. Russo}
\affil{Department of Astronomy, MS\,105-24, California Institute of Technology, Pasadena, CA 91125, USA}
\author{D. C. Homan, K. I. Kellermann, M. L. Lister}
\affil{National Radio Astronomy Observatory, 520 Edgemont Road, Charlottesville, VA 22903, USA}
\author{R. C. Vermeulen}
\affil{Netherlands Foundation for Research in Astronomy, P.O. Box 2, NL-7990 AA Dwingeloo, The Netherlands}
\author{E. Ros, J. A. Zensus}
\affil{Max-Planck-Instit\"ut f\"ur Radioastronomie, Auf dem H\"ugel 69, D-53121 Bonn, Germany}

\begin{abstract} 
We investigate the relation between the Doppler factor determined
from variations in total flux at 22 and 37\,GHz, and the apparent
transverse velocity determined from VLBA observations at 2\,cm. The
data are consistent with the relativistic beaming theory for compact
radio sources, in that the distribution of $\beta_{\rm app}/\delta_{\rm var}$,
for 30 quasars, is roughly consistent with a Monte Carlo simulation.
The intrinsic temperature appears to be $\sim 2 \times 10^{10}$\,K, close
to the ``equipartition value" calculated by Readhead (1994).  We deduce
the distribution of Lorentz factors for a group of 48 sources; the values
range up to about $\gamma=40$.

\end{abstract}

\section{Introduction}

Relativistic effects are commonly invoked to explain the superluminal
motion and high brightness temperature ($T_{\rm b}$) seen in compact radio
sources.  $T_{\rm b}$ can be found from VLBI measurements or from variations
in flux density; but these two methods depend differently on the Doppler
factor. By measuring $T_{\rm b}$ in both ways, the Doppler factor and the
Lorentz factor can be deduced, as well as $T_{\rm b}$(int), the intrinsic
temperature in the synchrotron source. L\"ahteenm\"aki et al.\ (1999a)
have done this for several sets of sources, and found that most of
the $T_{\rm b}$(int) have a range around $10^{11}$\,K. This is close to the
``equipartition'' value $\sim 5\times 10^{10}$\,K suggested by Readhead
(1994), and the diamagnetic limit $\sim 3\times10^{11}$\,K calculated
by Singal (1986). It is at the bottom end of the range suggested by
Kellermann \& Pauliny-Toth for the limit due to the inverse-Compton
catastrophe, $1-10\times 10^{11}$\,K. (See also Kellermann, 
these proceedings, page 185.)

In this paper we compare superluminal velocities, $\beta_{\rm app}$, with
$\delta_{\rm var}$, the Doppler factor derived from variability.  If a simple
relativistic beaming theory is correct, there will be a close connection
between these two quantities. The new large and reliable sample of
superluminal sources found in the VLBA 2\,cm survey (Zensus 
et al., 
these proceedings, page 27)
allows this to be done with some confidence, but see Section
6 for a discussion of the $\delta_{\rm var}$. We work with velocities
from the VLBA 2\,cm survey and Doppler factors from the Mets\"ahovi
flux density monitoring program at 22 and 37\,GHz (L\"ahteenm\"aki \&
Valtaoja 1999b, hereafter LV99). We convert the Mets\"ahovi values to
our assumed cosmology, $H_\circ=65$\,km\,s$^{-1}$\,Mpc$^{-1}$, 
$\Omega_{\rm m}=0.3$,
$\Omega_\Lambda=0.7$, but otherwise use the values found in their paper.

\section{The Velocities}

We only use velocities from the 2\,cm survey which have three or more
epochs of observation with an appropriately small error to the fit, and
which satisfy several criteria for morphology.  (Kellermann et al.\ 2003,
in preparation.)  The velocities are determined by a linear regression
through the locations of the components being followed, and the errors
shown in Figure~1 are the curve-fitting errors. We are using the 
\textsl{fastest}
component for each source, on the grounds that their velocities should
be representative of the true flow velocities. Slower-moving components,
especially those at a bend in the jet, may be dominated by backward
shock waves. Forward shock waves might also exist, and trying to
understand their role is also a goal of the survey. Other geometries
have been suggested for the jet, including a fast ``spine'' which we
would preferentially see, surrounded by a slower shell. In this case,
we think, the spine would also control the flux variations, so that
using the fastest (spine) velocity for $\beta_{\rm app}$ is appropriate. In
the next sections we assume that the pattern velocities are identical
to the flow velocities; we comment on this assumption in Section 7.

Five sources in common to the $\delta_{\rm var}$ and $\beta_{\rm app}$
lists were excluded on the grounds that they only had components situated
at a bend in the jet. We believe that in these cases we see a standing
shock wave, or perhaps a stationary location in a helical jet where the
flow is closest to the LOS, and hence boosted most strongly. In either
case the measured velocity is a poor indicator of the flow velocity,
and not useful in looking for relativistic effects. The final sample we
use contains 48 sources: 4 galaxies, 14 BL Lacs, and 30 quasars.


\section{The Doppler Factors}

The 22 and 37\,GHz light curves give the flux density in an outburst and
the time constant $\Delta t$. As a model we take $c\Delta t$ to be the
radius of an optically thick sphere. A solid angle is calculated and then
the brightness temperature $T_{\rm b}$.  (See L\"ahteenm\"aki et al.\ 1999a
for details.) A further assumption is now needed, that the measured
brightness temperature may be relativistically boosted above an intrinsic
temperature $T_{\rm b}$(int). LV99 choose $T_{\rm b}$(int)$=5\times 10^{10}$\,K, from
the estimates of equilibrium temperature for a self-absorbed synchrotron
source calculated by Readhead (1994). The variability Doppler factor is
calculated as $\delta_{\rm var} =$[$T_{\rm b}$(var)$/T_{\rm b}$(int)$]^{1/3}$.

A more direct way to get the brightness temperature of a compact source
is to measure its diameter and flux density with VLBI. The Doppler factor
can then be calculated by reference to an assumed $T_{\rm b}$(int). In most cases
of interest this yields a lower limit, because the cores of the sources
are only slightly resolved with terrestrial interferometer baselines
(Kellermann, 
these proceedings, page 185).

\begin{figure}[hbt!]  
\begin{center} 
\includegraphics[clip,width=0.9\textwidth]{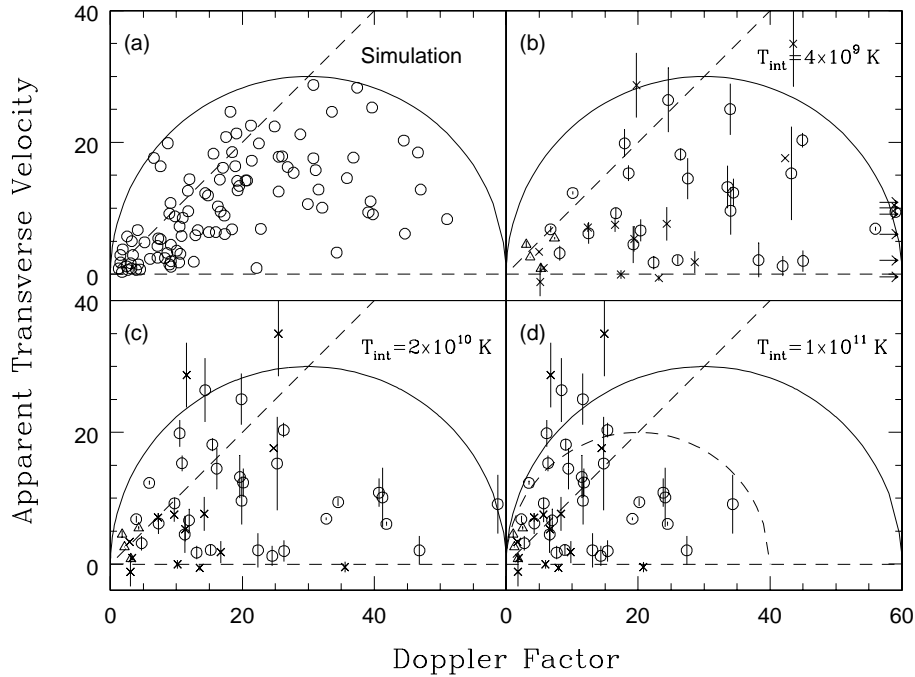} 
\end{center}
\vspace{-15pt} 
\caption {(a) Monte Carlo simulation of $(\beta_{\rm app}, \delta)$
for a flux-limited sample with N=100, $a=-1.25$.  The diagonal line
represents the ``$1/\gamma$ cone'', where the apparent velocity is
maximized for a fixed $\gamma$.  The semi-circle is the locus of
points for $\gamma=30$. (b) As in (a) with $\delta_{\rm var}$ calculated
for $T_{\rm int}=4\times 10^{9}$\,K.  Circles: Quasars; Crosses: BL Lacs;
Triangles: Galaxies. N=48 (c) As in (b) with $T_{\rm int}=2\times 10^{10}$\,K.
(d) As in (b) with $T_{\rm int}=1\times 10^{11}$\,K. $\gamma=20$ is shown
as the dashed curve.}

\end{figure}

An alternative method of finding a Doppler factor involves the inverse
Compton effect and the ratio of X-ray to radio flux. This method depends
on knowing the diameter from VLBI, but is only weakly dependent on the
X-ray flux, and so is close to the pure radio VLBI method described
in the previous paragraph.  A few cases have been studied in detail
and give good results (e.g., Unwin et al.\ 1994). 

\section{The $(\beta_{\rm app}, \delta)$ Relation}

Lister \& Marscher (1997, hereafter LM97) have made Monte Carlo
simulations of a flux-limited sample of compact sources drawn from a
population that has a distribution of Lorentz factor N($\gamma) \sim
\gamma^a$. They assume flux boosting of the form S$\sim \delta^2$,
which is appropriate for the flat-spectrum core.  Figure~1(a)
shows a realization for $a=-1.25$. (See LM97 for a discussion of the
exponents.) In this case the Lorentz factors are set such that $\gamma=30$
is the maximum value in the sample, so the points all fall under the
$\gamma=30$ curve.  Note that a majority of the sources lie under the
diagonal line $\beta_{\rm app}=\delta$, or $\theta \approx 1/\gamma$, as
they should, since the probability density for selecting a source peaks
near $\theta=1/2\gamma$ (Cohen 1989; Vermeulen \& Cohen 1994, hereafter
VC94; LM97). For a fixed $\gamma$, the apparent velocity is a maximum
on this line.

In Figures~1(b), (c) and (d) we show values of $(\beta_{\rm app},
\delta_{\rm var})$ for our sample, for three values of $T_{\rm b}$(int). We
believe that our data are representative of a flux-limited sample (Zensus
et al., 
these proceedings, page 27),
and LM97 show that $a=-1.25$ gives a reasonable
fit to the apparent--velocity distribution of the Caltech-Jodrell flat
spectrum survey (Taylor et al 1996).  Hence our points should lie within
a region bounded by a $\gamma=\rm{const}$ curve, as in Figure~1(a).
In Figure~1(b) 6 points are off scale on the x-axis, and the values
of $\delta_{\rm var}$ go to 100. In Figure~1(d) we show two $\gamma={\rm
const}$ curves, and it appears that, as in Figure~1(b), the points
will not properly fill any $\gamma$=const curve.  Figure~1(c) comes
closest to the simulation, and, given the small number of points (30),
it perhaps provides an adequate match.  These comments about Figure~1 are
confirmed by calculating the fraction $f$ of sources inside the $1/\gamma$
cone. For the Monte Carlo simulation $f=0.80$, and for the observations
the values for the 30 quasars are (for increasing temperature) 0.87,
0.77, and 0.50. On this basis $1\times 10^{11}$ can be excluded, and
$2\times 10^{10}$ is somewhat better than $4\times 10^{9}$.

\section{The distributions of $\beta_{\rm app}/\delta_{\rm var}$ and $\gamma$}

The product $\gamma\theta$ is a useful quantity because the probability
density function for $\theta$ depends only on $\gamma\theta$
for $\gamma^2\gg 1$ (see VC94, Fig.~7).  Hence the measured distribution of
$\gamma\theta$ can be compared with a theoretical distribution, or one
generated by a simulation, to see how closely the observations match
the standard theory.  In the previous section we found $f$, the fraction
of sources with $\gamma\theta <1$, and we picked $T_{\rm b}$(int) $= 2\times
10^{10}$\,K because its value of $f$ is close to the expected value.
However, it is clear that looking at many values of $\gamma\theta$
would be more powerful than looking at $f$ alone, as this would test
the general distribution throughout the region below the $\gamma=$
const curves.

\begin{figure}[hbt!]
\begin{center}
\includegraphics[clip,width=0.95\textwidth]{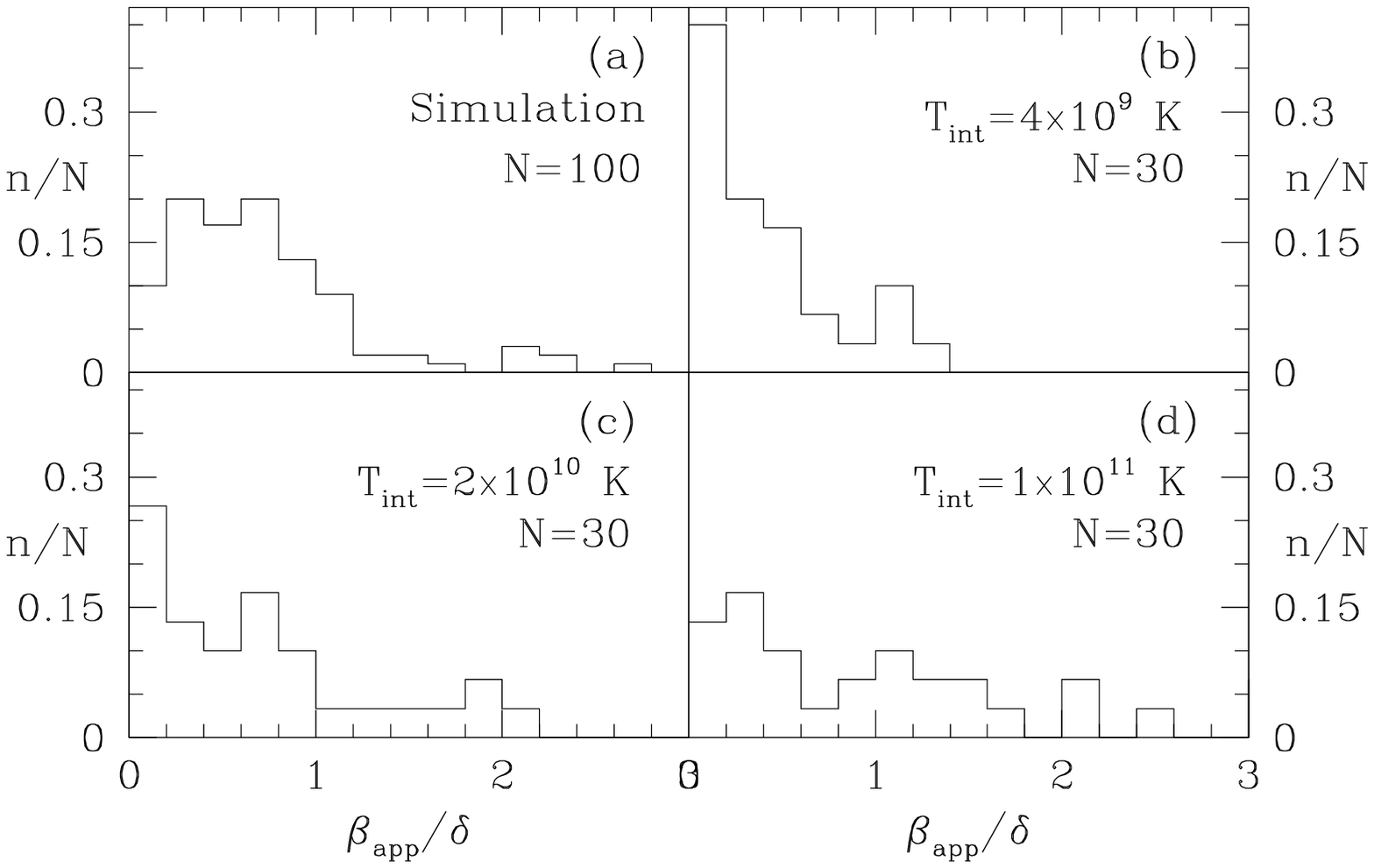}
\end{center}
\vspace{-15pt}

\caption
{Histograms of $\beta_{\rm app}/\delta = \gamma\theta$. (a) Results from
a Monte Carlo simulation, $N=100$. (b), (c), and (d) Calculated for
30 quasars, with three different values of $T_{\rm b}$(int) used to 
calculate $\delta_{\rm var}$.} 
\end{figure}

We actually use the ratio $\beta_{\rm app}/\delta_{\rm var}$, which, for
$\gamma^2\gg 1$, equals $\gamma\theta$. (A general result for a
relativistic beam is $\beta_{\rm app} = \beta\gamma\delta\,\sin\theta$, where
$\beta = v/c$.)  We form the histogram of $\beta_{\rm app}/\delta_{\rm var}$,
which counts the points between radii from the origin in Figure~1.
In principle we can mix galaxies, BL Lacs, and quasars even though they
have different parent populations, if we assume that $a$, and the slope
of the luminosity function, are the same for all the populations. These
assumptions seem somewhat dubious, however, and we only use the quasars.

Figure~2(a) shows the distribution of $\beta_{\rm app}/\delta_{\rm var}$ for the
100 sources in the simulation.  Figures~2(b), (c), and (d) are for the 30
quasars, for the three values of $T_{\rm b}$(int). The histogram in Figure~2(d)
is substantially flatter than that in 2(a) and supports our statement
that $T_{\rm b}$(int)$=1\times10^{11}$\,K is too high.  In Figure~2(b) the
distribution is too narrow, with a large spike in the first bin.  This
corresponds to the high values of $\delta_{\rm var}$ for this temperature. In
Figure~2(c) the histogram is a better match than the others, but it
still has an excess in the first bin.  We originally discarded 5 quasars
because they had slow or stationary components located at a bend, and we
believed that in those cases the measured velocity was not a good measure
of the flow velocity. If we had kept those sources, they would have been
in the first bin and the fit would be worse. In fact we discarded only
the most egregious cases, and it is likely that there are other sources
where we have measured a slow pattern rather than the flow velocity.
This, we believe, explains at least part of the excess in the first bin.

The number of sources is too small to allow us to pick a particular
``best'' value for $T_{\rm b}$(int). However, it  seems likely that for many
quasars the intrinsic temperature in the compact synchrotron source
is $1-3 \times 10^{10}$\,K.  For illustrative purposes below we use
$T_{\rm b}$(int)$=2\times10^{10}$\,K, and we apply it to galaxies and BL Lacs
as well as quasars.

We now calculate values of $\gamma$, using $T_{\rm b}$(int)$=2\times
10^{10}$\,K.  However, four BL Lacs have negative $\beta_{\rm app}$,
with values within 1.5$\sigma$ of zero. We assume that $\beta_{\rm app} +
2\sigma$ is an upper limit to the velocity, and calculate the resulting
upper limit to the Lorentz factor. The values of $\gamma$ found in this
way are nearly independent of how we pick $\beta_{\rm app}$, as can be seen
from Figure~1; the $\gamma=\rm{const}$ curves are vertical at $\beta_{\rm app}
= 0$, where $\gamma=\delta_{\rm var}/2$.  
We could instead adopt a lower
limit and assume that the negative sources are in the forward jet, but
moving towards the center.  In this case their flux  would be de-boosted,
whereas in fact they are strong sources.  Alternately, we could assume
that they are inward-moving in the back jet. We reject this because
there is plentiful evidence that in these compact sources we see only
forward jets. 

Figure~3 shows the distribution of Lorentz factor for the 48 sources, with
upper limits denoted by arrows.  The quasars have a broad distribution
between 5 and 25, with a few quasars both higher and lower.  The galaxies
and most of the BL Lacs have $\gamma<10$.  However, the two highest
$\gamma$ have large error bars (the two BL Lacs above the curves in
Fig.~1) and they are not reliable.  The quasars and BL Lacs have a
wide distribution of $\gamma$, but there are insufficient data to see
if they might match any of the distributions found in the Monte Carlo
simulations by LM97. We note that some of their distributions for $\gamma$
have a minimum in the first bin, as in Figure~3.  It is interesting
to note that Homan 
(these proceedings, page 35)
has estimated $\gamma$ for the
source \Index{3C\,279}, by analyzing a change in velocity, and assuming that it
is due to a change in $\theta$, not $\gamma$. He finds $\gamma\ga 15$.
Our calculation using $\delta_{\rm var}$ and $\beta_{\rm app}$ gives $\gamma=19$,
for $T_{\rm b}$(int)$=2\times 10^{10}$\,K. These methods of finding $\gamma$
are largely independent. Their good agreement adds to our conclusion that
for many sources the intrinsic temperature is $\sim 1-3\times10^{10}$\,K.

\begin{figure}[hbt!]
\begin{center}
\includegraphics[clip,width=0.9\textwidth]{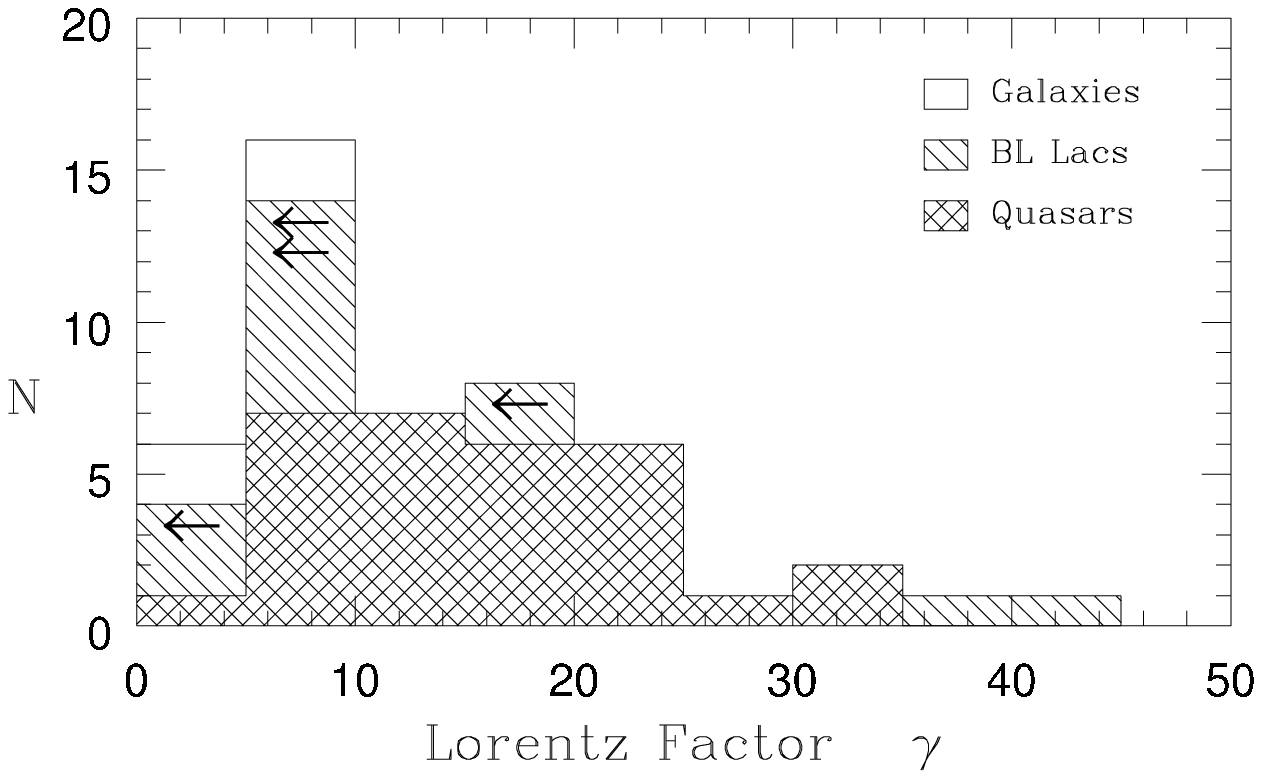}
\end{center}
\vspace{-15pt}
\caption{Histogram of Lorentz factors, calculated from 
$\beta_{\rm app}$ and $\delta_{\rm var}$ based on $T_{\rm int}=2\times 10^{10}$\,K.
Arrows indicate upper limits. $N=48$}
\end{figure}

\section{Reliability of $\delta_{\rm var}$}

We have several concerns over the reliability of the $\delta_{\rm var}$
values.  The first involves the procedure used to calculate $\delta_{\rm var}$
from a light curve. To partially address this we made some independent
determinations of $\delta_{\rm var}$ using the same procedures as
L\"ahteenm\"aki et al.\ (1999), and Valtaoja, et al. (1999).  We used the
8 and 15\,GHz variability curves from the University of Michigan web
site, and the 22 and 37\,GHz data from the Mets\"ahovi web site. We
found that the higher-frequencies gave better results, because there was
less blending between outbursts.  We did not try to investigate the effect
of non-overlap of the epochs of the $\delta_{\rm var}$ and $\beta_{\rm app}$
measurements; but we have seen that for most sources successsive
outbursts have a similar velocity.  Also, Valtaoja et al.\ (1999) state
that the apparent brightness temperature of well-defined flares does
not vary much in any one source.  However, the most important factor
appears to be the identification of individual outbursts, even at the
highest frequencies. This identification is somewhat subjective, and
apparently we were more conservative and our values for $\delta_{\rm var}$
were consistently about 30\% lower than the values published by LV99. In
a few cases there were larger discrepancies, apparently caused by
our missing some large flares not yet posted on the Mets\"ahovi web
site (Valtaoja, private communication). 
Thus we think that this systematic effect
might limit the accuracy of $\delta_{\rm var}$ to about 30\%.

The second general concern involves the model used for the source. The
calculation simply assumes that the time constant of the flare is
related to a radius by $r=c\Delta t$, and then that the solid angle
of the source is $\Omega=\pi(r/D)^2$, where $D$ is the metric distance
to the source. If we were to assume instead that the diameter, not the
radius, is given by $c\Delta t$, then the $T_{\rm b}$(int) would have to be
increased by a factor of 4 to reproduce the simulation in Figure~1(a).
Actually, $c\Delta t$ should be regarded as an upper limit to $r$; this
might further increase $T_{\rm b}$(int). Furthermore, to be more realistic,
the sphere should be replaced by a shock wave. See e.g., Marscher \& Gear
(1985).

\section{Pattern and Flow Velocities}

In this work we have assumed that the pattern velocity we measure
with the VLBA is the same as the flow velocity of the beam. However
it is clear that this is not always so. For example, we eliminated
several cases where the pattern is stationary at a bend in the jet,
on the grounds that the bright spot is due to a standing shock wave,
and its zero motion is not representative of the flow velocity.

VC94 made a simple model to study differing pattern and flow velocities;
namely, that the flow and the pattern have different $\gamma$,
and their ratio $r = \gamma_{\rm p}/\gamma_{\rm b}$ has a characteristic value.
From the early VLBI data they showed that if $r=1$ in all sources,
then $\gamma$ cannot be constant, and vice versa. From Figure~3 it is
quite clear that $\gamma$ is not constant in the compact sources. That
does not mean that $r$ is constant; indeed, since we count stationary
sources as having $r=0$, then $r$ has some distribution also.  If the
distribution of $r$ mostly has values below unity, then in Figure~1
$\beta_{\rm app}$ must be increased. The distribution of $\gamma$ will
shift to higher values, and its shape may change. It is difficult to
predict whether the best-matching $T_{\rm b}$(int) will increase or decrease,
as that depends on the details of the $r$ distribution.

\section{Conclusions}

We have combined variability Doppler factors with superluminal
motions, for a sample of about 50 radio sources. The distribution
of $\beta_{\rm app}/\delta_{\rm var}$ is roughly similar to that found in
a Monte Carlo simulation, provided the intrinsic temperature in the
synchrotron-emitting medium is $T_{\rm b}$(int)$ \sim 2\times 10^{10}$\,K.
This is near the ``equipartition temperature'', $\sim 5\times 10^{10}$\,K
suggested by Readhead (1994). It is below the upper limit based on the
inverse-Compton effect, $1-10\times 10^{11}$ (Kellermann \& Pauliny-Toth
1969; Kellermann, 
these proceedings, page 185),
and the diamagnetic limit calculated
by Singal (1986), $\sim 3\times 10^{11}$\,K.

The galaxies, and most of the BL Lacs, have $\gamma<10$, when calculated
with $T_{\rm b}$(int) $=2 \times 10^{10}$\,K. The quasars have a distribution
which is flat between $\gamma=5$ and $\gamma=25$, with only a few quasars
above and below these limits.


\begin{acknowledgements}
We are indebted to H. Aller, M. Aller, M. Kadler, and E. Valtaoja for
helpful discussions. The NRAO and the VLBA are operated by AUI, under
cooperative agreement with the NSF. This research has made use of data
from the University of Michigan Radio Astronomy Observatory which is
supported by funds from the University of Michigan.
\end{acknowledgements}

\printindex
\end{document}